\documentclass[aps,prl,superscriptaddress,showpacs,twocolumn]{revtex4}
\usepackage{graphicx}
\usepackage{color}
\usepackage{amssymb}

\begin{document}
\title{Nonlinear magnetic response in ruthenocuprates}

\author{I.\v Zivkovi\'c}

\affiliation{Institute of Physics, P.O.B.304, HR-10 000, Zagreb, Croatia}

\author{V.P.S. Awana}

\affiliation{National Physical Laboratory, Dr. K.S. Krishnan Marg, New Delhi-110012, India}

\author{H. Berger}

\affiliation{Institut de Physique de la Mati\`ere Complexe, EPFL, CH-1015 Lausanne, Switzerland}

\date{\today}
\begin{abstract}
We have performed an investigation of the nonlinear magnetic response in ruthenocuprates. A negative, diverging-like peak at the main magnetic transition $T_N$ in RuSr$_{2}$\emph{RE}Cu$_{2}$O$_{8}$ (\emph{RE} = Gd, Y) indicates a possible canted antiferromagnetic order. Another well defined feature above $T_N$ points to a blocking of superparamagnetic particles through the $T^{-3}$ dependence of the third harmonic at higher temperatures. Below $T_N$ a nondiverging peak appears, which is strongly affected by the addition of 10\% of Cu ions in the RuO$_2$ planes. In RuSr$_{2}$\emph{RE}$_{2-x}$Ce$_{x}$Cu$_{2}$O$_{10}$ the main magnetic transition $T_M$ is accompanied by two characteristic temperatures in the third harmonic of the ac susceptibility, in agreement with recent studies from $\mu $SR and M\"{o}ssbauer spectroscopy. We find that the spin-spin correlation temperature is the same in both families of ruthenocuprates.
\end{abstract}

\maketitle

\section{Introduction}

A possibility of coexistence of superconducting and ferromagnetic order on the microscopic scale has attracted a lot of attention to ruthenocuprates. Although a respectable amount of experimental work has been published so far, a complete and detailed description of the magnetic properties of the ruthenocuprates is still lacking. The ruthenocuprate family of materials consists of two well investigated phases, RuSr$_{2}$\emph{RE}Cu$_{2}$O$_{8}$ (Ru1212) and RuSr$_{2}$\emph{RE}$_{2-x}$Ce$_{x}$Cu$_{2}$O$_{10}$ (Ru1222) and RuSr$_{2}$RECe$_{2}$Cu$_{2}$O$_{12}$ (Ru1232) (RE = rare-earth) recently synthesized through a high-pressure-high-temperature (HPHT) procedure~\cite{Awana2005} with a composition RuSr$_{2}$RECe$_{2}$Cu$_{2}$O$_{12}$ (Ru1232) (RE = rare-earth). All ruthenocuprates have similar planar structure with RuO$_2$ planes responsible for the magnetic ordering and CuO$_2$ planes for the superconductivity. Between the two CuO$_2$ planes there is a \emph{RE} layer (as in YBa$_2$Cu$_3$O$_7$, where RuO$_2$ planes are replaced with CuO chains), a \emph{RE}$_{2-x}$Ce$_{x}$O$_2$ block or a \emph{RE}Ce$_2$O$_4$ block for Ru1212, Ru1222 and Ru1232, respectively. Due to the presence of the \emph{RE}$_{2-x}$Ce$_{x}$O$_2$ block, the Ru1222 system has adjacent Ru ions shifted along the $c$-axis by $(a+b)/2$ and the unit cell is doubled.

On general grounds, the Ru1212 system shows one magnetic transition around $T_N = 130$ K while the Ru1222 system is characterized by two: one around $T_M^F = 180$ K and the second one around $T_M = 100$ K (depends slightly on the \emph{RE}/Ce ratio). Although early reports have suggested a ferromagnetic order in Ru1212~\cite{Bernhard1999} and Ru1222~\cite{Felner1997}, neutron scattering results have indicated the presence of antiferromagnetism~\cite{Lynn2000,Jorgensen2001,McLaughlin2005} with an upper limit of $0.1 \mu _B$ for a ferromagnetic component. Moreover, the direction of the magnetic moment has been determined to be along the $c$-axis, contradicting the $\mu $SR~\cite{Bernhard1999}, EPR~\cite{Fainstein1999} and NMR~\cite{Tokunaga2001} measurements where it was concluded that the moments lie in the $ab$-plane.

To reconcile the proposed hypotheses, a weak ferromagnetism~\cite{Jorgensen2001,Felner2004}, a phase separation~\cite{Xue2003}, a combination of two~\cite{Felner2004a} and a spin-glass scenario~\cite{Cardoso2003} have been suggested. Recently, a $\mu $SR study~\cite{Shengelaya2004} on the Ru1222 system has shown that at $T_M^F$ only 15\% of the material gets ordered. This finding has been confirmed in a M\"{o}ssbauer study~\cite{Felner2004a}. The rest of the sample orders at $T_M$. As for the Ru1212 system, Xue and coworkers~\cite{Xue2003} have showed that substituting the Ru ions with the Cu ions leads to a separation of the ferromagnetic and the antiferromagnetic ordering temperatures. Moreover, a recent investigation of the nonlinear dynamics and the magnetization decay on the Ru1212 (\emph{RE} = Gd) composition~\cite{Cimberle2006} has revealed an existence of ferromagnetic clusters with an ordering temperature only few Kelvins above the $T_N$, the antiferromagnetic ordering temperature. A similar observation has been found on the Ru1212Y composition~\cite{Zivkovic2007}, although with a different sign of the third harmonic around $T_N$, which will be discussed later.

In this paper we extend our investigation of the nonlinear magnetic behavior in ruthenocuprates. We confirm our previous claim that the Ru1212 system, for \emph{RE} = Gd, Y, shows a negative third harmonic in the ac susceptibility which is not compatible with a simple AFM ordering of Ru ions. Through a detailed study of the ac field dependence of the ac susceptibility we show that at the main magnetic transition $T_N$ the third harmonic shows diverging-like behavior, where in simple AFM system, no divergence is observed both above and below $T_N$. Additional features are visible in both pure and doped Ru1212 systems which can be ascribed to a superparamagnetic behavior. The third harmonic for different \emph{RE}/Ce ratios in the Ru1222 system is qualitatively the same. Two characteristic temperatures around $T_M$ are found which indicates that a long-range ordering sets in at $T_M$.

\section{Experimental details}

We have performed the measurements on the following compositions: RuSr$_{2}$GdCu$_{2}$O$_{8}$ (Ru1212Gd), Ru$_{0.9}$Sr$_{2}$YCu$_{2.1}$O$_{7.9}$ (Ru1212Y) and RuSr$_{2}$\emph{RE}$_{2-x}$Ce$_{x}$Cu$_{2}$O$_{10}$ (Ru1222Eu) with $x$ ranging from 0.6 to 1.0. The polycrystalline samples used in this study have been measured previously, see Refs.~\cite{Felner1997,Awana2003,Zivkovic2002}. Ac susceptibility measurements were performed using the commercial CryoBIND system with the frequency of the driving field equal to 990 Hz.

Nonlinear susceptibilities can be defined through the expansion of the magnetization $M$ in the power series of the magnetic field $H$
\begin{equation}
M = M_0 + \chi _1 H + \chi _2 H^2 + \chi _3 H^3 + ...,
\end{equation}
where $\chi _1$ is the linear (or a first order) susceptibility and $\chi _2$ and $\chi _3$ are the second and the third order susceptibilities, respectively. Although usually much smaller then the linear component, $\chi _2$ and $\chi _3$ often provide additional information about the investigated system. It has been shown that the divergence of $\chi _3$ characterizes the spin-glass transition~\cite{Fujiki1981}. It has been used to distinguish spin-glasses from superparamagnets~\cite{Bitoh1993,Bitoh1996,Bajpai2001}, both showing similar behavior in the linear component $\chi _1$. For long-range-ordered systems $\chi _3$ shows divergence on both sides of $T_C$ in ferromagnets~\cite{Fujiki1981,Sato1981,Bitoh1993a,Nair2003}, while for antiferromagnets $\chi _3$ has a nondiverging, asymmetric shape at the transition with a positive sign of $\chi _3$ below $T_N$~\cite{Fujiki1981,Ramirez1992,Kushauer1995,Narita1996}.

Even order susceptibilities vanish when the magnetization has inversion symmetry with respect to the magnetic field applied. $\chi _2$ has been used to provide the evidence of the coexistence of the ferromagnetic and glassy behavior in reentrant spin-glass systems~\cite{Chakravarti1992} and doped cobalt-based perovskites~\cite{Androulakis2002}.

When an ac field $H = H_0 \cos \omega t$ with a frequency $\omega $ is applied, an induced voltage in the coils detected with a lock-in amplifier involves higher harmonic terms in addition to the first harmonic:
\begin{eqnarray}
\Delta V & \propto & - \frac{dM}{dt} \propto \omega H_0 [ \chi _1^{exp} \sin \omega t + \nonumber\\
& & + \chi _2^{exp} H_0 \sin 2\omega t + \frac{3}{4}\chi _3^{exp} H_0^2 \sin 3\omega t + \cdots ]
\end{eqnarray}
The harmonics are related to the higher-order susceptibilities through the following relations:
\begin{eqnarray}
\label{razvoj}
\chi _1^{exp} & = & \chi _1 + \frac{3}{4} \chi _3 H_0^2 + \frac{5}{8} \chi _5 H_0^4 + \dots , \nonumber\\
\chi _2^{exp} H_0 & = & \chi _2 H_0 + \chi _4 H_0^3 + \frac{15}{16} \chi _6 H_0^5 + \dots , \nonumber\\
\frac{3}{4} \chi _3^{exp} H_0^2 & = & \frac{3}{4} \chi _3 H_0^2 + \frac{15}{16} \chi _5 H_0^4 + \frac{63}{64} \chi _7 H_0^6 + \dots 
\end{eqnarray}
For small amplitudes $H_0$ we can put $\chi _1 = \chi _1^{exp}$, $\chi _2 = \chi _2^{exp}$ and $\chi _3 = \chi _3^{exp}$. There are no general rules as how large $H_0$ is allowed to be before higher order terms should be taken into account. The best option is to use as small $H_0$ as possible.

As explained above, the sign of the third harmonic is often more important in determining the qualitative aspects of the material under investigation then the absolute magnitude of the signal. Therefore, we have checked the sign of $\chi _3$ by measuring the triangular wave as an input signal for the lock-in amplifier~\cite{Sato1981}. In addition to that, the sign is also verified by observing the response from the superparamagnetic particles which should always be negative (see the following section).

Measurements for the ac field dependence of the third harmonic (Fig.~\ref{ac_field}) have been performed by measuring the signal in a temperature window around the peak since the maximum shift as the ac field is increased.

\section{Results and discussion}

\subsection{Ru1212}

\begin{figure}
\begin{center}
\includegraphics[width=8cm]{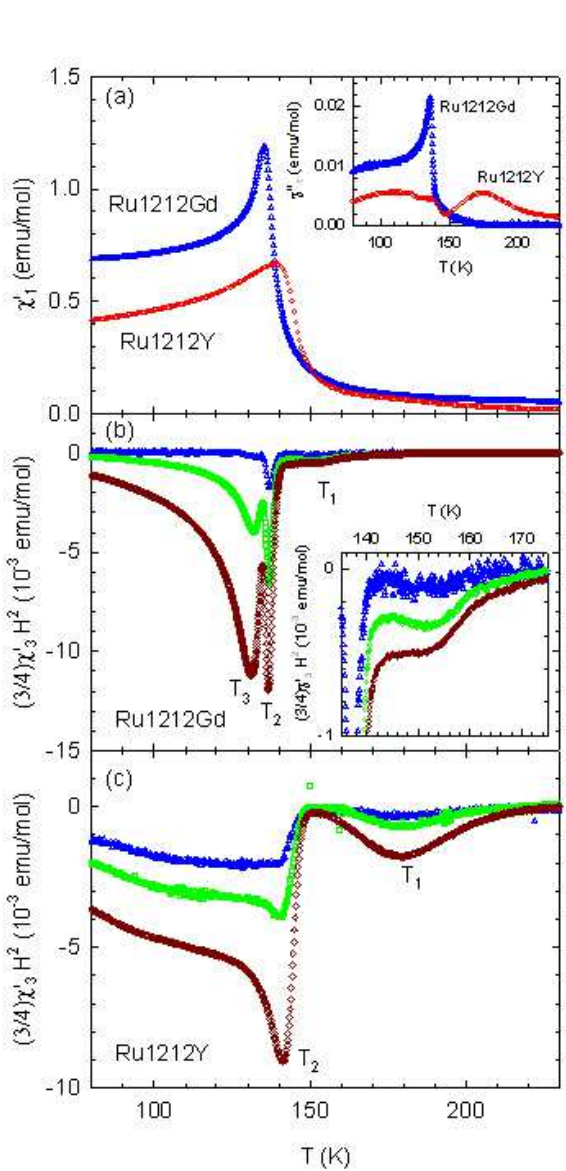}
\caption{(Color online) (a) Temperature dependence of the real part of the ac susceptibility for Ru1212Gd and Ru1212Y systems measured with $H_{AC} = 1$ Oe. The inset shows the imaginary part. (b) Temperature dependence of the third harmonic of the ac susceptibility for the Ru1212Gd system. From top to bottom $H_{AC} = 3, 10, 20$ Oe. Inset: enlarged view around $T_1$. (c) Temperature dependence of the third harmonic for the Ru1212Y system. From top to bottom $H_{AC} = 5, 10, 20$ Oe.}
\label{ru1212}
\end{center}
\end{figure}

We have investigated the Ru1212 system containing two rare-earth elements, Gd and Y. The real part of the ac susceptibility for the two systems is shown in Fig.~\ref{ru1212}a.
%
%
Ru1212Gd shows a somewhat larger susceptibility than Ru1212Y with a peak positioned at $T_N = 135$ K. For Ru1212Y the peak is more rounded with a maximum value around 140 K. The imaginary part of the ac susceptibility is displayed in the inset of Fig.~\ref{ru1212}a. A sharp peak is seen for Ru1212Gd system, while for Ru1212Y there are two broad maximums located around 110 K and 180 K, with a kink at 140 K where $\chi _1 '$ has a maximum.

Figure~\ref{ru1212}b shows the nonlinear susceptibility for Ru1212Gd measured in various ac fields. Three distinct features can be noticed, which occur at $T_1 = 152$ K, $T_2 = 137$ K and $T_3 = 131$ K. In the inset of Fig.~\ref{ru1212}b the peak at $T_1$ is shown enlarged. For small fields $T_1$ is barely visible and for larger fields it gets smeared out due to the growth of the peaks at $T_2$ and $T_3$. The peak at $T_2$ is very sharp for all the fields applied but it overlaps with the feature at $T_3$ as the field increases. $T_3$ peak shows a rapid growth as the field is increases, with a long tail below the temperature where the minimum occurs.

In Fig.~\ref{ru1212}c the third harmonic for the Ru1212Y system is presented. A broad maximum around 180 K has been attributed to the formation of the superparamagnetic particles~\cite{Zivkovic2007}. This feature leaves a visible imprint in the first harmonic as well (Fig.~\ref{ru1212}a). On the other hand, in Ru1212Gd neither $\chi _1'$ nor $\chi _1''$ show visible deviation at $T_1$. Below 150 K $\chi _3'$ starts to grow (in negative values) with a kink at 140 K (where a maximum in $\chi _1'$ is located) for small fields. As the field is increased, a peak develops at a temperature slightly above 140 K with a tail for lower T. No $T_3$ peak is observed in the Ru1212Y system. Taking into account that the RuO$_2$ planes in the Ru1212Y system investigated in this paper are slightly disordered due to the doping with extra Cu ions, it is naturally to conclude that the features observed at $T_1$ and $T_3$ are intrinsic to Ru1212 ruthenocuprate (as seen for the Ru1212Gd composition).

\begin{figure}
\begin{center}
\includegraphics[width=8cm]{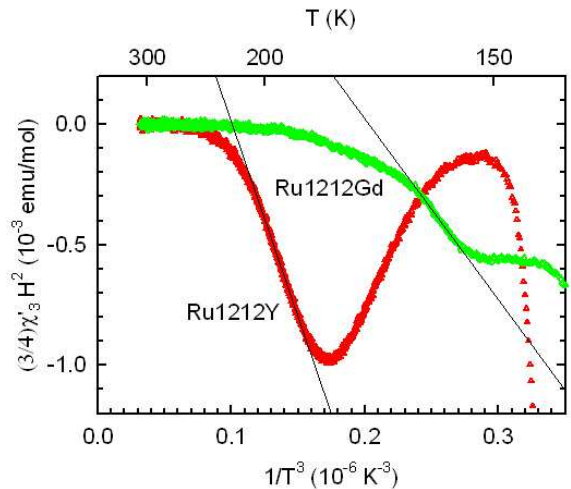}
\caption{(Color online) $T^{-3}$ dependence of $\chi _3'$ for Ru1212Y and Ru1212Gd measured in 10 Oe and 20 Oe, respectively. Solid lines represent the best fit to Wohlfarth's model (see text).}
\label{T-3}
\end{center}
\end{figure}

It has been shown~\cite{Bitoh1996} that the existence of the superparamagnetic particles can be verified through the $T^{-3}$ dependence of $\chi _3'$. According to the Wohlfarth's superparamagnetic blocking model~\cite{Wohlfarth1979}, $\chi _1'$ of the assembly of superparamagnetic particles follows a Curie law above the blocking temperature $T_B$ while $\chi _3'$ shows negative $T^{-3}$ dependence,
\begin{eqnarray}
\label{Tminus1}
\chi _1' & = & \frac{n\left\langle \mu  \right\rangle }{3} \frac{\left\langle \mu  \right\rangle }{k_B T} \\
\label{Tminus3}
\chi _3' & = & - \frac{n\left\langle \mu  \right\rangle }{45} \left ( \frac{\left\langle \mu  \right\rangle }{k_B T} \right )^3,
\end{eqnarray}
where $n$ is the number of particles per unit volume, $\left\langle \mu  \right\rangle $ is the average magnetic moment of the single particle and $k_B$ is the Boltzmann constant. In Fig.~\ref{T-3} we show appropriate plots for Ru1212Gd and Ru1212Y.
%
%
The linear dependence of $\chi _3'$ on $T^{-3}$ is found only in a small temperature interval: between 188 K and 203 K for Ru1212Y and between 156 K and 161 K for Ru1212Gd. Although in conventional superparamagnetic systems the particle's internal spin-spin correlation temperature is much higher than the blocking temperature $T_B$~\cite{Bitoh1993,Luo1991}, it has been shown recently~\cite{Bajpai2000} that for Li$_{0.5}$Ni$_{0.5}$O a similar behavior occurs with a 10 K wide temperature interval where the third harmonic is linear in $T^{-3}$. We are aware that a 5 K interval observed in the Ru1212 system is probably too small and it can serve only as an indication. What is important is that a doped system Ru1212Y shows qualitatively the same behavior as a pure system (\emph{RE} = Gd) corroborating the hypothesis that these features are intrinsic to material and that they are not the consequence of the existence of impurities.

From equations~(\ref{Tminus1}) and~(\ref{Tminus3}) one can extract the average magnetic moment of the superparamagnetic particle~\cite{Bitoh1993}. However, due to the presence of the ordering at $T_N$ and the paramagnetic contribution from the Gd ions (in the Ru1212Gd system), it was in this case not possible to extract the magnetic moment.

It is indicative that for the Ru1212Y composition, for which 10\% of Ru ions have been replaced by Cu ions, $T_1$ peak is larger then in the stoichiometric Ru1212Gd composition and has shifted to higher temperatures. The same happens for the first harmonic of Ru1212Y (see Fig.~\ref{ru1212}). This suggests that the incorporation of Cu ions in the RuO$_2$ plane enhances the formation of superparamagnetic particles. It doesn't affect the main transition since $T_2$ shows the same behavior for both compositions.

$\chi _3'$ gradually vanishes at higher temperatures ($\approx 250 - 280$ K). This applies for both investigated Ru1212 systems, indicating a common mechanism behind the build-up of correlations.

In both Ru1212\emph{RE} systems (\emph{RE} = Gd, Y) investigated here, the third harmonic remains negative, contrary to the report of Cimberle and coworkers~\cite{Cimberle2006}. In Ref.~\cite{Cimberle2006} two positive peaks have been observed with the interpretation that there are one positive and one negative peak, the negative one hollowing the positive peak. The positive peak has been ascribed to an AFM order while the negative peak has been attributed to the blocking of the superparamagnetic particles~\cite{Cimberle2006}. The close inspection reveals that the only difference between Ref.~\cite{Cimberle2006} and our results lies in the sign of the third harmonic, since the two peaks from Ref.~\cite{Cimberle2006} are the $T_2$ and $T_3$ peaks from Fig.~\ref{ru1212}b. 

%
\begin{figure}
\begin{center}
\includegraphics[width=8cm]{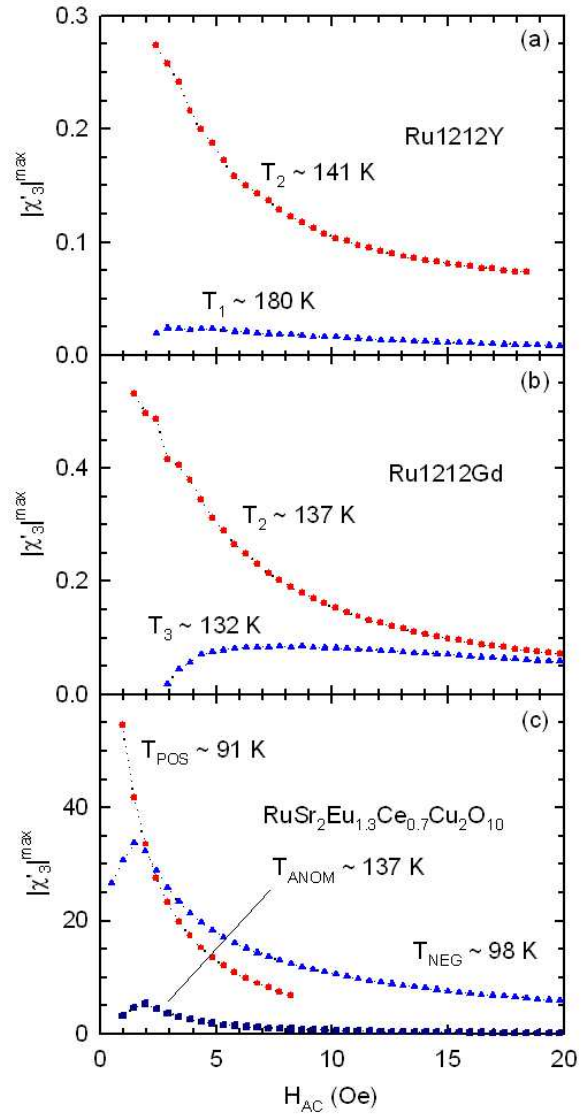}
\caption{(Color online) ac field dependence of the amplitude of the peaks in $\chi _3$ for various ruthenocuprates. The peaks shift in temperature as the ac field is increased and the labels correspond to the low-field value. Dotted lines are guides for the eye.}
\label{ac_field}
\end{center}
\end{figure}

The detailed ac field dependence (Fig.~\ref{ac_field}a,~\ref{ac_field}b) shows that the $T_2$ and $T_3$ peaks have substantially different behavior in the small field regime, which has not been probed in Ref.~\cite{Cimberle2006}. $T_2$ shows a divergent-like behavior, while $T_3$ starts to decrease below $\approx 8$ Oe and is not observable below 2 Oe. Similar ac field dependence has been observed in Ref.~\cite{Cimberle2006} (for fields above 7 Oe), except for the sign, but interpreted as a negative peak hollowing the positive one. We have checked our setup applying the triangular wave to the lock-in amplifier to confirm the correct sign of the third harmonic. In addition, the existence of $T_1$ in the Ru1212 system with a $T^{-3}$ dependence indicates an occurrence of superparamagnetic particles. There is no doubt that this results in a \emph{negative} $\chi _3$ (eq.~(\ref{Tminus3})), as reproduced in our measurements. $T_1$ is a natural explanation for the occurrence of blocked superparamagnetic particles which gives rise to time relaxations of magnetization~\cite{Cimberle2006}. We may assume that the phase of the third harmonic in Ref.~\cite{Cimberle2006} was simply changed by 180 degrees, either before the measurement or during the data analysis.

The main magnetic transition in the Ru1212 system is characterized by a negative, diverging peak at $T_2$ for both compositions investigate. Due to the presence of the adjacent peaks and relatively small signal, we were unable to perform the critical analysis which would allow us to determine to which class this transition belongs. Divergence in the limit of $H_{AC} \rightarrow 0$ indicates a long-range ordered magnetic state. This line of reasoning has been used before to discriminate between a spin-glass and a superparamagnetic system~\cite{Bajpai1997,Bajpai2001}. Also, investigation of ferromagnets in the limit of $H_{AC} \rightarrow 0$ showed~\cite{Nair2003} divergence on both sides of $T_C$. The negative character of $T_2$ peak is in disagreement with the proposed C-type AFM system~\cite{Lynn2000} for which it is expected to show a positive, nondiverging third harmonic for $T < T_N$ and vanishingly small signal for $T > T_N$~\cite{Fujiki1981,Narita1996}. On the other hand, it has been shown~\cite{Narita1996} that canted AFM systems diverge negatively on both sides of the transition. In addition, another well defined peak has been observed in~\cite{Narita1996} below the transition, which has been ascribed to an interaction of domain walls with an external field. We also see the appearance of a peak at $T_3$ for larger fields. Based on this experimental evidence, we suggest a canted AFM ordering to occur in the Ru1212 system.

Canted AFM structure has been previously proposed for the Ru1212 system~\cite{Jorgensen2001}. Calculations of the local spin-density approximation of Nakamura and Freeman~\cite{Nakamura2002} showed that canted AFM has a slightly lower energy then c-type ordering seen in the neutron scattering. Investigation of ac susceptibility in dc-bias field~\cite{Mandal2002} revealed a metamagnetic transition which was suggested to be between the canted AFM state for fields below the critical field and FM state above.

Since an upper limit for a ferromagnetic component at 0.1 $\mu _B$ has been obtained~\cite{Lynn2000}, it was hard to accommodate large canting angles to explain three times larger magnetic moment revealed from the magnetization measurements~\cite{Butera2001}. Xue and coworkers~\cite{Xue2003} suggested a phase separation into an AFM matrix and FM particles which eliminates the need for a large canting angles of the AFM matrix. This scenario is also supported by our measurements.

The largest difference between the two Ru1212 compositions investigated in this work is revealed below the main transition. In the Gd-based compound there is another well-defined peak $T_3$ with a strong ac field dependence while the Y-based compound shows only a broad feature with a modest field dependence. $T_3$ peak shows nondiverging behavior while the broad feature in Ru1212Y is visible even for smallest measuring fields. Very similar observation has been reported in the case of a canted AFM system~\cite{Narita1996}. The appearance of a peak below the main transition has been attributed to the effect of the external field on magnetic domains formed by weak ferromagnetic moments. Taking into account the fact that the $T_3$ peak is missing in Ru1212Y where Cu ions to some extent alter the genuine magnetic order in RuO$_2$ planes, we conclude that it is intrinsic to the magnetic order in Ru1212.

\subsection{Ru1222}

\begin{figure*}
\begin{center}
\includegraphics[width=16cm]{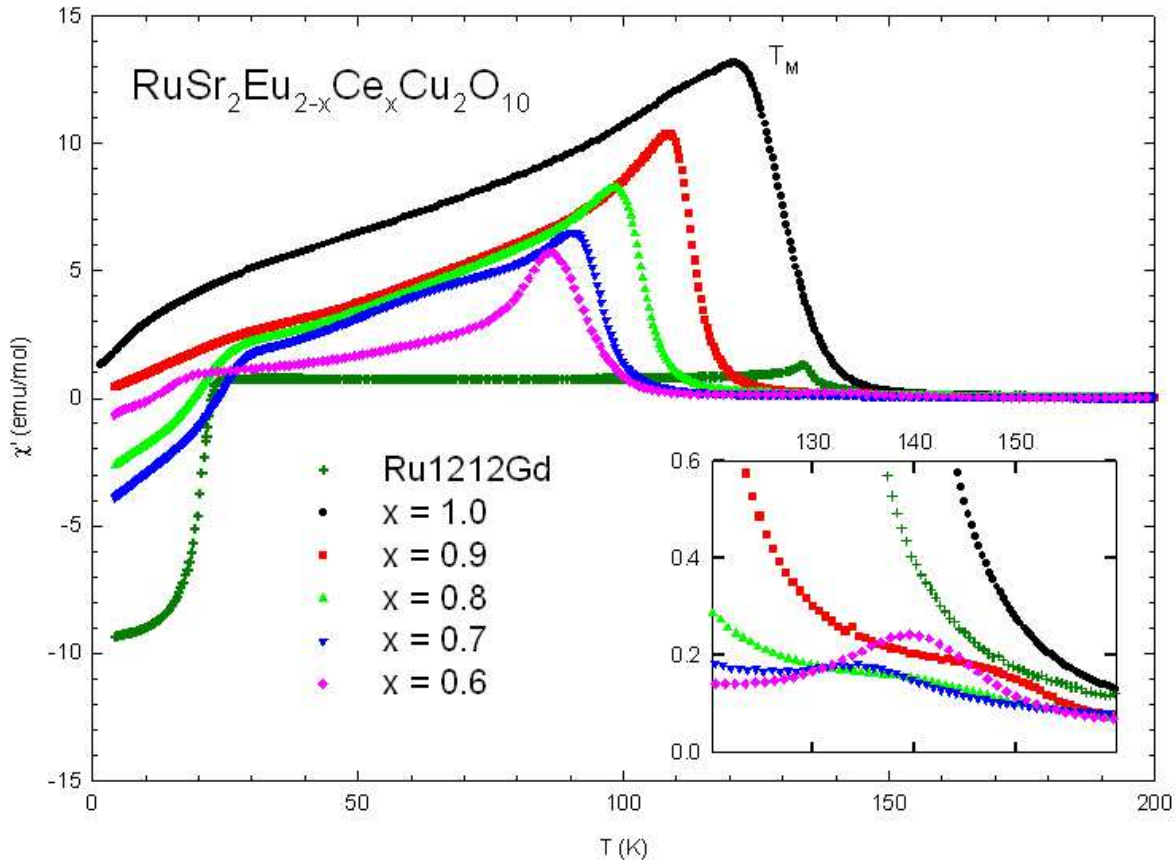}
\caption{(Color online) The first harmonic for Ru1222Eu and Ru1212Gd compositions. Inset: enlarged temperature window where the anomaly for the Ru1222 system occurs. No systematic behavior has been observed, either in the temperature or in the size of the anomaly.}
\label{svi-wide}
\end{center}
\end{figure*}

The first harmonic in the Ru1222Eu system for the concentrations ranging from $x = 1.0$ to $x = 0.6$, along with Ru1212Gd data, is presented in Fig.~\ref{svi-wide}.
%
%
In general, susceptibility of the Ru1222 system is approximately an order of magnitude larger than for the Ru1212 system. As $x$ decreases both the temperature $T_M$ and the size of the peak decrease, from 121 K for $x = 1.0$ to 85 K for $x = 0.6$. For concentrations with $x \leq 0.8$ there is a kink in around 30 K indicating an onset of the superconductivity. Similar results have been obtained through the dc susceptibility measurements~\cite{Felner2002}.

In addition, there is an anomaly in the susceptibility curve which occurs between 120 K $< T_{ANOM} < 140$ K, shown in the inset of Fig.~\ref{svi-wide}. No correlation has been observed between either the size of the anomaly or the temperature at which anomaly occurs with respect to the nominal Eu/Ce ratio, in agreement with~\cite{Felner2005}. In $\mu $SR study~\cite{Shengelaya2004}, conducted on the $x = 0.6$ composition, it was concluded that the anomaly does not represent a bulk transition. Although all the compositions were prepared at the same time and in the same laboratory, $x = 0.6$ shows particularly strong signal. Also, an overall shape for this concentration is somewhat different from other curves. On the other hand, $x = 1.0$ composition shows monotonic behavior without the apparent anomaly. We will show below that the anomaly is also present for this concentration but can be only observed in the third harmonic, while in the first harmonic it overlaps with the peak at $T_M$.

\begin{figure}
\begin{center}
\includegraphics[width=8cm]{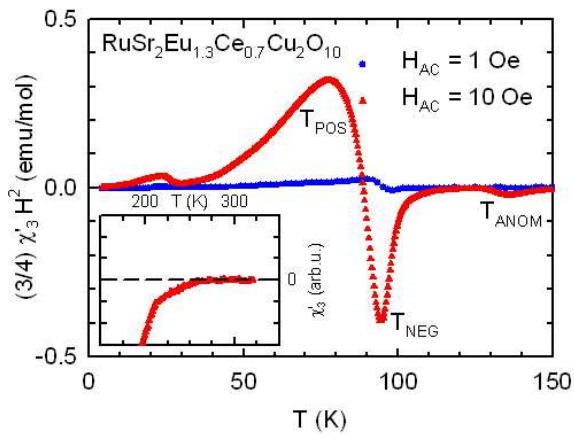}
\caption{(Color online) The third harmonic for the Ru1222Eu system with $x = 0.7$ in two different ac fields. The high temperature part is shown in the inset.}
\label{ru1222eu13}
\end{center}
\end{figure}

The third harmonic for the RuSr$_{2}$Eu$_{1.3}$Ce$_{0.7}$Cu$_{2}$O$_{10}$ ($x = 0.7$) composition measured in 1 and 10 Oe is shown in Fig.~\ref{ru1222eu13}. Three distinct magnetic features can be discerned in larger fields: a small negative peak around the temperature where the anomaly in $\chi _1$ has been observed ($T_{ANOM}$), a negative peak above $T_M$ and a positive peak below $T_M$. On lowering the temperature the signal becomes smaller, until the superconducting order sets in below 30 K. The third harmonic measurements have been recently used to prove the coexistence of ferromagnetic and superconducting order parameters~\cite{Leviev2004}. For other concentrations the results are very similar, with $T_{POS}$ and $T_{NEG}$ shifting in temperature according to the shift in $T_M$. In the inset there is an enlarged view of the high temperature part for $H_{AC} = 10$ Oe where we show the dissapearence of the third harmonic in the same temperature range as for the Ru1212 system (Fig.~\ref{T-3}).
%
%

\begin{figure}
\begin{center}
\includegraphics[width=12cm]{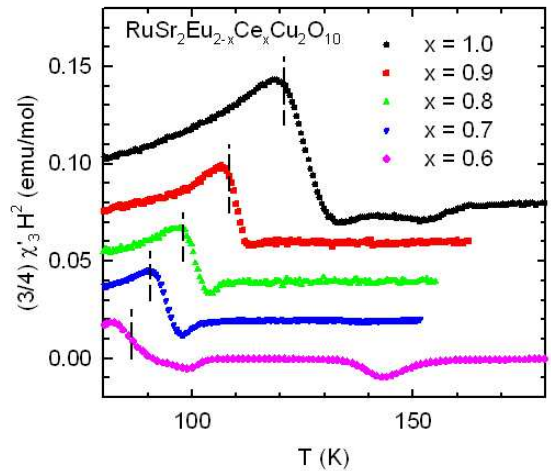}
\caption{(Color online) The third harmonic for the Ru1222Eu system measured with 1 Oe. The curves have been vertically displaced for clarity. The vertical dashed lines mark $T_M$, the maximum in $\chi _1$.}
\label{ru1222-1Oe}
\end{center}
\end{figure}

As we have mentioned in the introduction, it is very important to measure the higher order harmonics in as small a field as possible, to be able to use the approximation $\chi _3 = \chi _3^{exp}$ (see eq.~(\ref{razvoj})). In Fig.~\ref{ru1222-1Oe} we show all the investigated concentrations of the Ru1222Eu system measured in 1 Oe. All the curves show a similar pattern: a positive peak below $T_M$ (vertical dashed lines) and a small negative dip above $T_M$. This is a strong indication that $T_{POS}$ and $T_{NEG}$ are related to the main magnetic transition $T_M$. For $x = 0.6$ and $x = 1.0$ $T_{ANOM}$ is already observed for $H_{AC} = 1$ Oe.

The occurrence of two characteristic temperatures around the main magnetic transition in the Ru1222 system has been reported in recent investigations by $\mu $SR~\cite{Shengelaya2004} and M\"{o}ssbauer spectroscopy~\cite{Felner2004a}. These reports showed the existence of two internal magnetic fields appearing around the main magnetic transition $T_M$. In addition, the existence of the ordering just above $T_M$ has been indicated in our previous reports~\cite{Zivkovic2002a,Zivkovic2006}. A temperature dependence of the time relaxations of the ac susceptibility~\cite{Zivkovic2002a} and the peculiar inverted hysteresis~\cite{Zivkovic2006} has shown that these phenomena originate at a slightly higher temperature than $T_M$ and fully develop below $T_M$.

With respect to the occurrence of the anomaly, all compositions show the same behavior, even the parent compound ($x = 1.0$). This is important to stress since a recent report~\cite{Felner2005} indicated that $x = 1.0$ does not show the anomaly. We suggest that the anomaly is not visible in the first harmonic of the ac susceptibility and magnetization due to the higher magnetic transition at $T_M$ for $x = 1.0$. It has been suggested that for $x < 1.0$ the reduction of the Ce content leads to oxygen depletion. Ru$^{5+}$ ions surrounded by oxygen holes reduce to Ru$^{4+}$, which has been assumed to be related to the occurrence of the anomaly. Our measurements show that if the clustering of Ru$^{4+}$ ions is related to this feature, it is not the Ce content that drives the reduction from Ru$^{5+}$ to Ru$^{4+}$ ions, since the RuSr$_{2}$EuCeCu$_{2}$O$_{10}$ ($x = 1.0$) compound is stoichiometric. NMR experiments on Ru1212~\cite{Tokunaga2001}, which is also stoichiometric, indicated the coexistence of Ru$^{5+}$ and Ru$^{4+}$ ions which has been associated with the transfer of electrons from CuO$_2$ to RuO$_2$ planes and the occurrence of superconductivity in this compound. We propose that a similar mechanism might also be present in the Ru1222 system. The transfer of electrons must be weaker than in the Ru1212 system, since $x = 1.0$ and $x = 0.9$ are not superconducting. With further substitution of Eu$^{3+}$ for Ce$^{4+}$ ions additional holes are introduced to CuO$_2$ planes which induces superconductivity for $x \leq 0.8$.

It is instructive to look at the ac field dependence of characteristic peaks for the Ru1222 system, $T_{POS}$, $T_{NEG}$ and $T_{ANOM}$. We have only shown the results for the Ru1222Eu $x = 0.7$ composition but other compositions, including $x = 1.0$, reveal qualitatively similar results. Fig.~\ref{ac_field}c shows that $T_{POS}$ has a diverging character, while $T_{NEG}$ and $T_{ANOM}$ are nondiverging, although this is only evident in fields smaller then 2 Oe, which indicates the importance of the small-field measurements. A very similar observation in an amorphous ferromagnet, Fe$_5$Co$_{50}$Ni$_{17-x}$Cr$_x$B$_{16}$Si$_{12}$ with $x$ = 5~\cite{Nair2003}, has been explained invoking a clusterization above the main transition, before the full FM order sets in. This resulted in a negative, nondiverging third harmonic above $T_C$ and a positive, diverging harmonic below $T_C$. We propose that a similar situation occurs in the Ru1222 system. As shown by $\mu $SR study~\cite{Shengelaya2004}, just above $T_M$ a majority of the volume orders and from our results it seems clear that there is no long-range order. Eventually at $T_M$, where the rest of the sample gets ordered~\cite{Shengelaya2004}, $\chi _3$ diverges indicating a long-range order. This is to be contrasted with recent neutron scattering results on Ru1222, $x = 0.8$ where no long-range order could be measured for the main magnetic phase~\cite{Lynn2007}. Moreover, it is claimed~\cite{Lynn2007} that the ordering at $T_M$ is actually related to the impurity phase of unknown origin and that Ru ions incorporated in the Ru1222 phase do not contribute significantly to the observed magnetic behavior in the Ru1222 system. The systematic change of characteristic temperatures with $x$ in linear and nonlinear magnetic dynamics presented in this report and in previous investigations, with various dopants for Ru ions~\cite{Felner2005} suggest an intrinsic scenario behind the magnetism in the Ru1222 system. More experiments are needed in order to elucidate the microscopic nature of magnetic ordering in this material.
%
%

\begin{figure}
\begin{center}
\includegraphics[width=12cm]{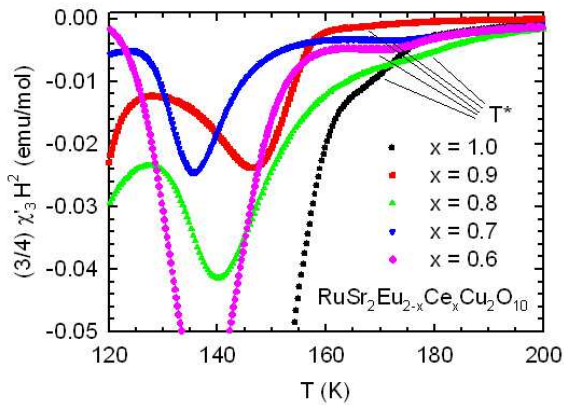}
\caption{(Color online) Measurement of the third harmonic with $H_{AC} = 10$ Oe for all the investigated concentrations of the Ru1222Eu system. $T_{ANOM}$ is developed for all the concentrations. $x = 1.0$ has $T_{ANOM}$ and $T_{NEG}$ overlapped.}
\label{above_anomaly}
\end{center}
\end{figure}

Some features are not visible in small fields ($\sim 1$ Oe). We show in Fig.~\ref{above_anomaly} measurements performed with $H_{AC} = 10$ Oe. The anomaly is now clearly visible for all the concentrations, with the $x = 0.6$ concentration showing the largest signal. For the $x = 1.0$ composition $T_{ANOM}$ is overlaps with a large, negative peak at $T_{NEG}$. Above $T_{ANOM}$ there is another deviation, 170 K $< T^{\ast } < 180$ K, which for some concentrations develops into a peak and for others creates only a barely visible shoulder. The most pronounced peak is again seen for the $x = 0.6$ composition. Around the same temperature a formation of superparamagnetic clusters has been proposed from nonlinear magnetization measurements~\cite{Xue2003} and $\mu $SR~\cite{Shengelaya2004} showed the existence of magnetic order in 15\% of the sample below $T^{\ast }$. The weak, negative sign of the third harmonic supports the hypothesis of a minor volume fraction ordering locally and giving rise to magnetism above the main magnetic transition $T_M$. Due to the small signal and large background from other peaks, we were unable to find an appropriate temperature interval with a $T^{-3}$ dependence, as we have shown for the Ru1212 system (see Fig.~\ref{T-3}). The Wohlfarth's model, which predicts a $T^{-3}$ dependence, assumes a constant average moment of the particle (see eqs.~(\ref{Tminus1}) and (\ref{Tminus3})). As suggested in Ref.~\cite{Garcia2006}, due to the different temperature dependence of the FM and AFM interactions inside the clusters, the average magnetic moment is temperature dependent. This implies a nondiverging third harmonic with a temperature dependent slope in the $\chi _3$ versus $T^{-3}$ plot, as in our case.

The origin of the high-temperature ordering continues to be a subject of debate. Except for the obvious intrinsic scenario, an impurity-based explanation has been proposed~\cite{Felner2005} with Sr-Ru-Cu-O$_3$ phase showing similar temperature dependence of the coercive field as the Ru1222 system. Cu$^{2+}$ ions are thought to be inhomogeneously distributed in both Ru and Sr sites which causes Sr-Ru-Cu-O$_3$ phase to act as an independent particle inside the Ru$^{5+}$ matrix. This study has been conducted on a system with a long-range ferromagnetic order where magnetic domains and domain walls play a dominant role in the mechanism behind the coercivity. On the other hand, nanosized particles incorporated in the Ru1222 matrix can be considered as monodomain structures, with a superparamagnetic blocking as the main mechanism generating the coercive field. Although an anisotropy ($K$) is involved in both processes, a temperature dependence of the coercive field in a bulk system should not be taken as an indicator for nanosized particles. This leaves the intrinsic scenario as a probable mechanism, but we are still missing the microscopic explanation of it.

Other features observed in this report are also unlikely to be related to the presence of impurities. The Ru1212 system has been investigated before~\cite{Cimberle2006} and, apart from the sign of the third harmonic, the same features have been observed. Furthermore, through the investigation of Ru$_{0.9}$Sr$_{2}$YCu$_{2.1}$O$_{7.9}$ (Ru1212Y) we have shown that upon introduction of a structural disorder due to the incorporation of Cu ions into the RuO$_2$ planes, the Ru1212 system's predominant ordering at $T_2$ does not change. On the other hand, the peak at $T_2$, which presumably reflects the interaction of domain walls in the canted AFM state with the external field, is strongly influenced with the imposed disorder, indicating that the magnetic response from the Ru1212Gd compound is intrinsic to the Ru1212 system.

%
%

In the Ru1222 system all the different \emph{RE}/Ce ratios show consistent behavior between the first and the third harmonic. $T_{POS}$ and $T_{NEG}$ change in accordance with the change in $T_M$ and in the $\mu $SR experiment~\cite{Shengelaya2004} it has been shown that this involves more than 90\% of the sample's volume. The detection limit of x-ray diffraction measurements for our samples indicates $< 3$\% of impurities~\cite{Felner1997,Awana2003,Zivkovic2002}, confirming that $T_{POS}$ and $T_{NEG}$ are intrinsic to the Ru1222 system. The anomaly around 120 K has been observed in all previously investigated Ru1222 samples and has been linked to the high temperature transition around 180 K~\cite{Felner2005,Garcia2006}. Substitution of Ru ions with Mo ions~\cite{Felner2005} showed that while the main magnetic transition is shifted, the anomaly remains unchanged. If the anomaly is related to some sort of impurities in the Ru1222 system, one would expect drastic changes in position and intensity, which has not been observed. In addition, the higher harmonics are orders of magnitude weaker than the first harmonic and we have shown that the anomaly appears even for the smallest fields used. This strongly implies that the anomaly is intrinsic to the Ru1222 system.

\section{Conclusion}

Several novel features have been observed in our study of the nonlinear susceptibility of ruthenocuprates. In Ru1212 we have found a negative third harmonic of the ac susceptibility, with a clear separation between the main magnetic transition and the formation of superparamagnetic particles. The divergent-like behavior of the third harmonic at the main magnetic transition indicates a long-range ordered state. Previous reports favored a canonical AFM state with magnetic moments pointing along the c-axis. Our results contradict this hypothesis since canonical AFM systems are expected to show a nondiverging positive third harmonic. We propose that the majority of magnetic moments order in a canted AFM state, in accordance with the neutron diffraction results. The dominant ferromagnetic response comes from the separated, short-range ordered particles which are blocked below a temperature slightly higher than the temperature of the main magnetic transition. The peak appearing below $T_N$ for larger magnetic fields has been ascribed to an interaction between domains of weak ferromagnetic moments and the applied field.

Nonlinear response in the Ru1222 system revealed two characteristic temperatures around $T_M$, in line with $\mu $SR and M\"{o}ssbauer results. The charateristic lower temperature $T_{POS}$ coincides with the main magnetic transition $T_M$ seen in the linear response. The divergence of the third harmonic at $T_{POS}$ is an indication of the onset of a long-range order. We have observed a small negative feature in $\chi _3$ around 180 K for all compositions. This is a possible signature of a minority phase ordering into supeparamagnetic particles.

In both ruthenocuprate systems the third harmonic starts to show in the temperature range $\approx 250 - 280$ K. Below this temperature short-range correlations bind individual spins into larger clusters giving rise to an observed nonlinearity. It is important to notice that when the third harmonic starts to show a deviation from the Curie-Weiss behavior occurs, which can be crucial when determining a paramagnetic moment from the Curie-Weiss plot.



\end{document}